\documentclass[a4paper]{jpconf}
\usepackage{graphicx}
\usepackage{psfrag,multirow}
\usepackage{amsmath, amsthm, amssymb}
\begin{document}
\title{Phase transitions and non-equilibrium relaxation in kinetic models of opinion formation}

\author{Soumyajyoti Biswas$^1$, Anjan Kumar Chandra$^1$, Arnab Chatterjee$^{2,3}$, Bikas K. Chakrabarti$^{1,4,5}$}

\address{$^1$Theoretical Condensed Matter Physics Division,
Saha Institute of Nuclear Physics, 1/AF Bidhannagar, Kolkata 700064, India.}

\address{$^2$CMSP Section, The Abdus Salam International Centre for Theoretical Physics,
Strada Costiera 11, Trieste I-34014 Italy.}

\address{$^3$Centre de Physique Th\'{e}orique (CNRS UMR 6207), Universit\'{e} de la M\'{e}diterran\'{e}e Aix Marseille II,
Luminy, 13288 Marseille cedex 9, France.}

\address{$^4$Centre for Applied Mathematics and Computational Science,
Saha Institute of Nuclear Physics, 1/AF Bidhannagar, Kolkata 700064. India.}

\address{$^5$Economic Research Unit, Indian Statistical Institute,
203 Barrackpore Trunk Road, Kolkata 700108, India.}

\ead{soumyajyoti.biswas@saha.ac.in, anjan.chandra@saha.ac.in, arnab.chatterjee@cpt.univ-mrs.fr, bikask.chakrabarti@saha.ac.in}

\begin{abstract}
\noindent We review in details some recently proposed kinetic models of opinion dynamics.
We discuss  several variants including a generalised model. 
We provide mean field estimates for the critical points, which are numerically supported 
with reasonable accuracy. Using non-equilibrium relaxation techniques, we also investigate 
the nature of phase transitions observed in these models.
We also study the nature of correlations as the critical points are approached.
\end{abstract}
\section{Introduction}
\label{sec:Introduction}
Application of statistical physics to understand social dynamics is an interesting
and very active area of research~\cite{ESTP,Stauffer,Castellano,Galam:1982,Galam:2008}.
The key question one asks is how a set of interacting individuals choose between
different options (vote, language, culture, opinions etc), leading
to a state of `consensus' in one such option, or a state of coexistence of
many of them.
Opinion dynamics is one of the most important aspects of a society. It is
a collective dynamical phenomena. Numerous models have been introduced so far
to study the dynamics that leads to different opinion states and the processes
that determine transitions between such states. Voter model which has a binary
opinion variable with the opinion alignment proceeding by a random choice of
neighbours ~\cite{Holley} and the Sznajd discrete opinion formation 
model where more than just a pair of spins is associated with the decision 
making procedure ~\cite{Sznajd} are two such examples. In some other models more than 
two 
opinions and also opinion as a continuous variable has been considered 
\cite{Hegselmann,Deffuant,Fortunato,BCS}. Models where the range of interactions are more than
nearest neighbors but finite and even time dependent (see Ref~\cite{soham} and references therein) are also studied.
 `Opinions' are subject to changes due to 
binary or group interactions, global feedback and even external factors.
The usual interest in these studies lies in the distinct steady state properties:
a phase characterized by individuals with widely different opinions and 
another phase with a measured fraction of individuals with similar opinions.

We focus our attention to a specific class of models proposed 
recently~\cite{Lallouache:2010a,Lallouache:2010b},
having apparent similarity with kinetic models of wealth exchange~\cite{CC-CCM}.
The opinions of individuals are continuous variables in $[-1,1]$ which change 
due to
binary interactions. The only parameter in these models is `conviction', which 
is a measure
of how much an individual sticks to his/her previous opinion while interacting 
with another.
The system of such individuals, or the `society', reaches a state of 
`consensus' if this parameter stays above a threshold. One can also generalise 
this model~\cite{Sen:2010}
by introducing another parameter modeling the `influence' of the other 
individual. This study models the fact that the ability to influence need not 
be identical to one's conviction.
When `influence' and `conviction' are identical, one gets the original 
model~\cite{Lallouache:2010a,Lallouache:2010b}. The case of `consensus' is an ordered phase 
(full or partial) while another `disordered' phase exists, characterized by 
null (or very small) value of individual opinions.

In this article, we will first review the original model and its generalisation. 
Then we will present some interesting features of the above model~\cite{Lallouache:2010a,Lallouache:2010b}, 
and propose some simplifications and variants. In 
particular, using Non-Equilibrium Relaxation (NER)~\cite{Hinrichsen:2000}, 
we study the phase transition observed in these models.
The associated critical exponents are obtained. Further, we propose a simpler variant of the original model to 
capture the minimum ingredients required to obtain the phase transition. Also 
the effect of global feedback is studied.
The paper is organized as follows:
In Sec.~\ref{sec:lccc} and Sec.~\ref{sec:2} we will discuss the 
original model and two of its variants.
In Sec.~\ref{sec:3}, we discuss the general results, and
the non-equilibrium relaxation in Sec.~\ref{sec:ner}.
Next, we discuss 
a generalized model in Sec.~\ref{sec:Sen}, 
a map in Sec.~\ref{sec:map} and also
introduce a new model in Sec.~\ref{sec:new}.
We conclude with a summary and discussions of our main results in 
Sec.~\ref{sec:sum}.

\section{The LCCC model}
\label{sec:lccc}
The basic idea of this model originated from a multi-agent statistical model
of closed economy \cite{CC-CCM} where $N$ agents exchange a quantity $w$
defined as wealth. Initially each agent begin with a certain amount of 
wealth $w_i$, $i = 1,2, \ldots, N$ such that $w_i > 0$ and the
total wealth $W = \sum_i w_i $ is conserved. The system evolves with a
prescribed trading rule where agents interact with each other through a
pairwise interaction characterised by a ``saving" parameter $\lambda$, with
$0 \le \lambda \le 1$. The dynamics  (CC model) is as follows :
\begin{eqnarray}
w_i^{\prime} &=& \lambda w_i + \epsilon (1 - \lambda) (w_i + w_j)\nonumber\\
w_j^{\prime} &=& \lambda w_j + (1 - \epsilon) (1 - \lambda) (w_i + w_j)
\end{eqnarray}
where $\epsilon$ is a stochastic variable that changes with time and as $w$ is
a conserved quantity, for each transaction $w_i^{\prime} + w_j^{\prime} = w_i 
+ w_j$, where $w_i^{\prime}$ and $w_j^{\prime}$ are the agent wealth after
the transaction. The functional form for steady state distribution $f(w)$
is seen to be close to the $\Gamma$ distribution~\cite{Patriarca}. 

Following this, Lallouache et. al.~\cite{Lallouache:2010a,Lallouache:2010b} 
proposed a minimal multiagent model for the collective dynamics of opinion  formation. 
Let $o_i(t) \in [-1,+1]$ be the opinion of an individual $i$ at 
time $t$. In a system of $N$ individuals, 
opinions change out of binary interactions:
\begin{eqnarray}
\label{eq:lccc}
 o_i(t+1) &=& \lambda[o_i(t) + \epsilon o_j(t)] \nonumber \\
 o_j(t+1) &=& \lambda[o_j(t) + \epsilon^\prime o_i(t)]
\end{eqnarray}
   where $\epsilon$, $\epsilon^\prime$ are drawn randomly from uniform 
distributions in $[0,1]$. Here, $\lambda$ is a parameter,
which is interpreted as `conviction'. The above model~\cite{Lallouache:2010a,Lallouache:2010b} 
(LCCC model hereafter) 
considers a society where everyone has the same value of conviction $\lambda$. 
It is important to note that there are no conservation laws here. 
The opinions are bounded, i.e., $-1 \le o_i(t) \le 1$. The ordering in the 
system is measured by 
a quantity (order parameter) $O= |\sum_i o_i |/N$. Another important quantity is the so called 
`condensation fraction' $p$, which is the fraction of the agent having $o_i=\pm 1$.  
Numerical simulations show that the multiagent system (dynamics given by 
Eqn.~(\ref{eq:lccc})) goes into either of 
the two possible phases: for any $\lambda \le \lambda_c$, $o_i=0$ $\forall i$, while for 
$\lambda > \lambda_c$,
$O>0$ and $O \to 1$ as $\lambda \to 1$, with $\lambda_c \approx 2/3$. 
$\lambda_c$ is the critical point of the phase transition.
The relaxation time, defined as the time to reach a stationary value of $O$ in 
time, diverges as $\tau \sim |\lambda-\lambda_c|^{-z}$ when 
$\lambda \to \lambda_c$ on either side. A
similar behavior is also observed for $p$. Although the values of the 
exponents differ 
($z\approx1.0\pm0.1$ and $z\approx 0.7\pm0.1$ with $O$ and $p$ respectively, 
reported in~\cite{Lallouache:2010b}), the critical points are same.
The order parameter exponent $\beta$ is defined (for any order parameter $X$) 
as:
\begin{equation}
\label{eq:beta}
X \sim (\lambda - \lambda_c)^{\beta}.
\end{equation}
For $O$ its value is $0.10 \pm 0.01$. Now if one also attempts to fit the growth of $p$ in a similar form, one finds its value to be $0.95 \pm 0.02$ 
.

A mean field calculation can be proposed for the fixed point $o^*$:
\begin{equation}
 \label{eq:lcccmf}
o^*[1 - \lambda (1 + \langle \epsilon \rangle)]=0,
\end{equation}
from which it is easy to show that the critical point is $\lambda_c = 1 /(1+ \langle \epsilon \rangle)$
($\langle \ldots \rangle$ refers to average).
For uniform random distribution of $\epsilon$, $\langle \epsilon \rangle = 1/2$ and hence,
$\lambda_c = 2/3$. It is important to note that this mean-field treatment does not
incorporate the cut-offs at $\pm 1$ and yet gives the correct critical point.
We also note that the underlying topology ($1d$, $2d$ or infinite range) 
has barely any effect on the critical
point.

It is appropriate to mention here that a map version of the 
model~\cite{Lallouache:2010a,Lallouache:2010b} has also been proposed:
\begin{equation}
 \label{eq:maplccc}
o(t+1) = \lambda (1+\epsilon(t)) o(t)
\end{equation}
where $\epsilon(t)$ is drawn randomly from a uniform distribution in $[0,1]$,
and $o(t)$ is bounded in $[-1,+1]$ as before. The critical value $\lambda_c$
can be analytically shown to be $\exp[-(2\ln 2 -1)] \approx 0.6796$~\cite{Lallouache:2010b}. 

\section{Two more models}
\label{sec:2}
\subsection{A simpler model}
\label{sec:c}
\begin{figure}
\includegraphics[width=16.0cm]{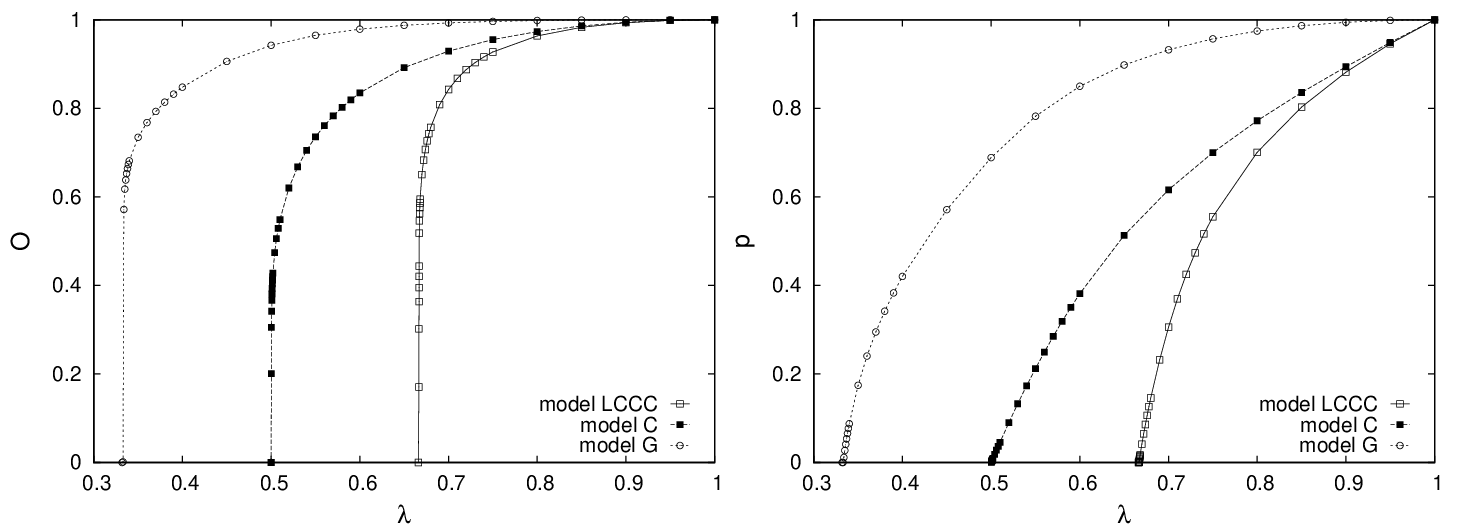}
   \caption{The phase diagrams for the 3 models.
\textit{Left}: Behavior of order parameter $O$.
\textit{Right}: Behavior of condensation fraction $p$.}
\label{fig:allop}
\end{figure}

A simpler model (model C hereafter) can be proposed:
\begin{eqnarray}
\label{eq:c}
 o_i(t+1) &=& \lambda o_i(t) + \epsilon o_j(t) \nonumber \\
 o_j(t+1) &=& \lambda o_j(t) + \epsilon^\prime o_i(t)
\end{eqnarray}
   where $\epsilon$, $\epsilon^\prime$ are drawn randomly from an uniform distribution in $[0,1]$ as earlier. 
Here, $\lambda$ is a parameter as in LCCC model, which is interpreted as `conviction'. 
An individual $i$ upon meeting another individual $j$, retains his own opinion 
proportional to his conviction but also picks up a random influence of that of the other.
Numerical simulation reveals that the system goes into either of 
two possible phases: for any $\lambda \le \lambda_c$, $O_i=0$ $\forall i$, a \textit{symmetric} phase, while for $\lambda > \lambda_c$,
$O>0$ and goes to $1$ as $\lambda \to 1$, a \textit{symmetry broken} phase, with $\lambda_c \approx 1/2$. $\lambda_c$ is the
critical point of the phase transition (Fig.~\ref{fig:allop}).

A mean field calculation similar to the one discussed earlier, can be proposed for the fixed point $o^*$:
\begin{equation}
 \label{eq:cmf}
o^*(1 - \lambda - \langle \epsilon \rangle)=0,
\end{equation}
from which it is easy to show that the critical point is $\lambda_c = 1 - \langle \epsilon \rangle$.
For uniform random distribution of $\epsilon$, $\langle \epsilon \rangle = 1/2$ and hence,
$\lambda_c = 1/2$.
\subsection{Global effect on LCCC model}
\label{sec:g}
In the context of social opinion formation, global opinion often takes a vital
role in influencing one's opinion. In that case, a person $i$, apart
from being ``influenced" stochastically by a person $j$, is also ``influenced"
stochastically by the average opinion of the entire society at that moment.
Mathematically the dynamics can be represented by
\begin{eqnarray}
\label{eq:global_lccc}
 o_i(t+1) & = & \lambda[o_i(t) + \epsilon o_j(t)] + \epsilon^{\prime} O(t) \nonumber \\
o_j(t+1) & = & \lambda[o_j(t) + \eta o_i(t)] + \eta^{\prime} O(t),
\end{eqnarray}
where $\epsilon$, $\epsilon^{\prime}$, $\eta$ and $\eta^{\prime}$ are random 
numbers, drawn from uniform distribution in $[0,1]$.
In this case (model G hereafter), the \textit{symmetry broken} phase $O \ne 0$
appears for $\lambda > 1/3$, and for $\lambda \le 1/3$ the system is in a
\textit{symmetric} phase, with $O_i=0$ $\forall i$ and all individual agents have the opinion $0$
(Fig.~\ref{fig:allop}).

This transition point can again be explained by a mean-field approach.
At the steady state, i.e. when $O$ reaches a steady value, Eqn.~(\ref{eq:global_lccc})
can be written as,
\begin{equation}
\label{eq:mfglccc}
 o^{*} = \lambda(1 + \langle \epsilon \rangle)o^{*} + \langle \epsilon^{\prime} \rangle o^{*}
\end{equation}
from which it can be easily shown that $\lambda_c = 1/3$.

As in LCCC, the critical points of these two models have barely any
effect due to change in the underlying geometry. 
In what follows we study the multi-agent dynamics of these  models  and their non-equilibrium relaxation.
\section{Results}
\label{sec:3}

As mentioned above for the mean-field version of LCCC, 
the order parameter exponent is, $0.10 \pm 0.01$.
The order parameter exponent  for the model C is found to be 
$0.17 \pm 0.01$ for $O$ and again if one fits the behavior of $p$ in a similar form it comes out to be $0.98 \pm 0.02$, which is similar to the values found for LCCC.
For the model G, we estimate the exponent $\beta = 0.081 \pm 0.001$ and for $p$ it is
$0.85 \pm 0.01$.

We also calculate the relaxation behavior of the order parameters. An order parameter 
$X$ relaxes to equilibrium in time $t$ as 
\begin{equation*}
X(t) \propto \exp(-t/\tau) 
\end{equation*}
for the symmetric phase (going to $0$) and
\begin{equation*}
X(t) \propto X_0 [1 - \exp(-t/\tau)] 
\end{equation*}
in the symmetry-broken phase, $X_0$ being the equilibrium value of $X$. Away from the critical point,
we plot this relaxation behavior for different values of the parameter $\lambda$, to extract the value of $\tau$.
We observe that the relaxation time diverges as $\tau \sim |\lambda - \lambda_c|^{-z}$ both below and above the critical
point $\lambda_c$, with roughly the same exponent $z$.
\begin{figure}
\begin{center}
        \psfrag{ylabel}{$\tau$}
        \psfrag{xlabel}{$ ( \lambda-\lambda_c) $}
        \psfrag{insetylabel}{$ \tau $}
        \psfrag{insetxlabel}{$ | \lambda-\lambda_c| $}
            \includegraphics[width=8.5cm]
        {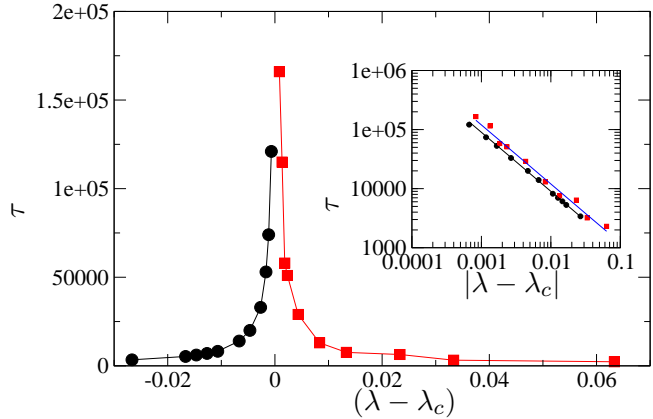}
    \end{center}
\caption{Behavior of relaxation time $ \tau $ with $ \lambda- \lambda_c$ from the order parameter $O$, 
for model LCCC (From Ref.~\cite{Lallouache:2010b}). The system size was $N = 500$.}
\label{fig:relaxmap}
\end{figure}
For LCCC (mean field), 
our estimate for  $z$ is $0.97 \pm0.01$ for $O$ while it is $1.10 \pm 0.01$ for $p$, which is consistent
with another estimate ($1.16 \pm0.03$ in Ref.~\cite{Sen:2010}) but is different from that reported in Ref.~\cite{Lallouache:2010b}.
For model C,  our estimates for  $z$ are $1.58 \pm0.01$ for $O$ and $1.34 \pm 0.01$ for $p$.
In case of model G, for $O$, $z = 1.2\pm0.1$ and for $p$, $z = 1.75\pm0.01$.

In the following subsection, we also report the study of a lattice version of the LCCC model. 
In this version, the agents are arranged on a $1d$ lattice, 
and a randomly chosen nearest neighbor pair update their
opinions according to Eqn.~(\ref{eq:lccc}). The critical behavior of this model is studied in detail.

\section{Non-equilibrium relaxation}
\label{sec:ner}
\begin{figure}
\centering \includegraphics[width=8.0cm]{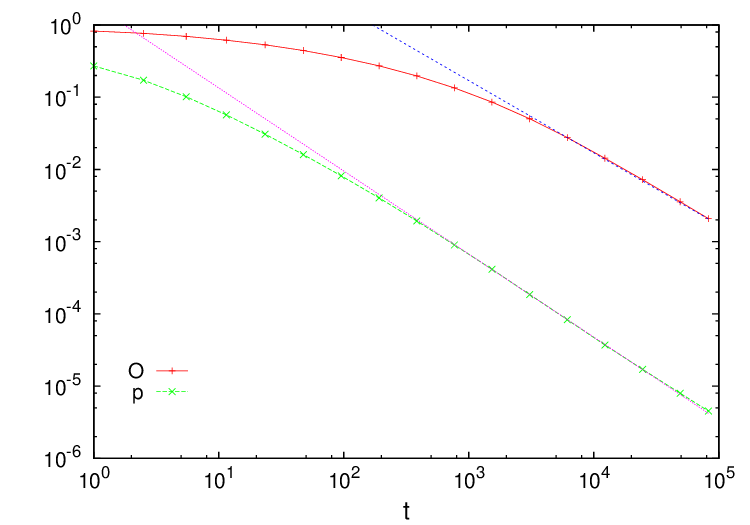}
\caption{The relaxations of the two quantities $O$ (order parameter) and $p$ (condensation fraction) are shown at the critical point 
($\lambda_c = 0.66679 \pm 0.00001$) for the $1d$ version of LCCC model. 
The power-law decays fit to exponent $1.00\pm0.05$ for $O$ and $1.15\pm0.01$ for $p$. The system size was $N=1200$.}
\label{delta}
\end{figure}
NER is a well established simulation strategy to investigate the phase transitions 
of systems in the thermodynamic limit (for a review see Ref.~\cite{Hinrichsen:2000,Ozeki:2007}). 
In this method the system is allowed to relax from an initial non-equilibrium state to the 
equilibrium state. It turns out that with much smaller amount of systematic errors, 
the critical exponents can be estimated from this method. 

All thermodynamic quantities show a relaxation in time, 
starting with initial values away from equilibrium. 
This temporal relaxation of the NER functions 
(above mentioned thermodynamic quantities) is used to determine the critical point 
accurately, as well as the critical exponents. 
The most important quantity in this study is of course the relaxation of 
the order parameter for the 
transition. 
We study the relaxation of both $O$ and $p$  at and near the critical point.

Here we first study the lattice version of the LCCC model in details, 
later we get back to the infinite range model as well.
The dynamics is started from an initial condition with full order 
(i.e., all agents have same extreme opinion, $+1$ or $-1$). 
Away from criticality, the system is expected to relax exponentially to its equilibrium order parameter value. 
At the critical point, however, the order parameter will relax asymptotically following a power law of the form
\begin{equation}
O(t)\sim t^{-\delta}.
\label{eq:delta}
\end{equation}
Note that a similar power-law  is also  observed for the condensation fraction $p$.
Fig.~\ref{delta} shows the 
behavior of the two quantities at the critical point. From the slope of the log-log plot, the critical exponent $\delta$ can be found.
From this figure we observe that the critical point for both the quantities are same, and also the critical exponent $\delta$ is
very close for both these quantities ($\delta=1.15\pm0.01$ for relaxation of $p$ and $1.00\pm0.05$ for $O$).

\begin{figure}
\includegraphics[width=8.0cm]{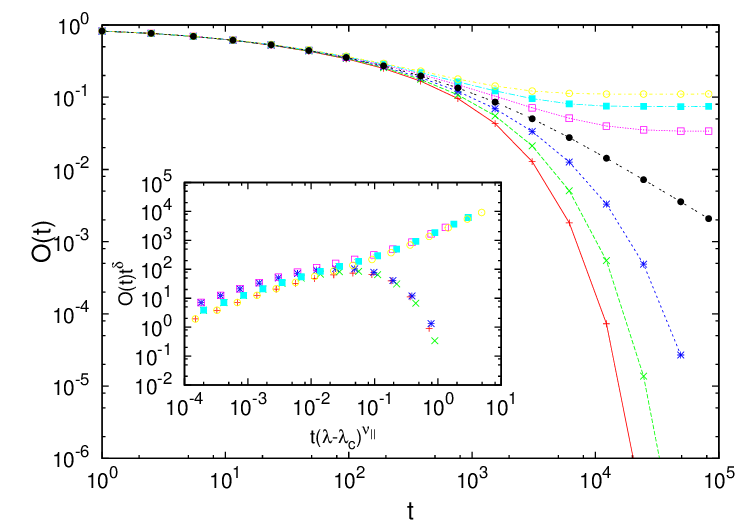}
\includegraphics[width=8.0cm]{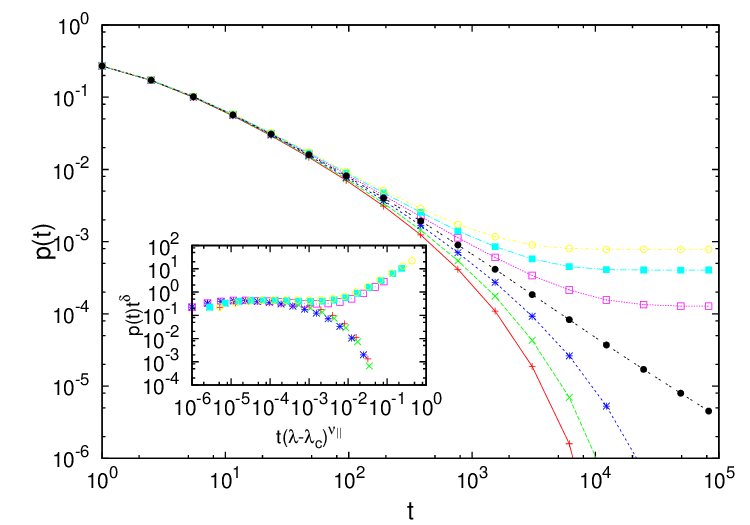}
\caption{\textit{Left}: Relaxation of the order parameter $O$ near $\lambda_c$ for the $1d$ version of LCCC model. 
The parameter $\lambda$ differs by $0.00001$ for the successive sets. The power law is obtained (central curve) for $\lambda_c=0.66679$. 
The system size for the simulation is $N=1200$. 
Inset: Plot of  $t(\lambda-\lambda_c)^{\nu_{||}}$  against $O(t)t^{\delta}$ using the  previously obtained 
value of $\delta$. The data collapse is obtained for  $\nu_{||} \approx 1.2 \pm 0.1$.
\textit{Right}: Same for the condensation fraction $p$.
Inset: Plot of $t(\lambda-\lambda_c)^{\nu_{||}}$ against $p(t)t^{\delta}$ using the previously obtained value of $\delta$. 
The data collapse is obtained for $\nu_{||} \approx 1.5 \pm 0.1$.
}
\label{nu-collapse-o}
\end{figure}

The critical point of the transition can be obtained with high accuracy by plotting the decay of the order parameter at 
and on either side of the critical point. Fig.~{\ref{nu-collapse-o}} shows the variation of the order parameter $O$ 
for different $\lambda$ values near criticality.
The accurately estimated critical point turns out to be $\lambda_c=0.66679\pm0.00001$ for $N=1200$, 
which is very close to the value ($2/3$) quoted in \cite{Lallouache:2010b,Sen:2010}. 
A similar estimate was made with $p$ (Fig.~{\ref{nu-collapse-o}}), the estimate of the critical point turns out to be exactly the same.

\begin{figure}
\centering \includegraphics[width=8.0cm]{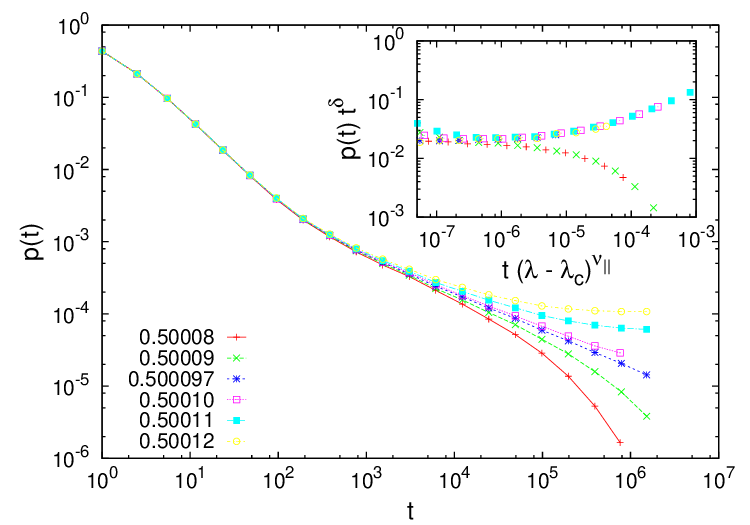}
   \caption{Relaxation of the condensation fraction $p$ near $\lambda_c$ for model C. 
The system size for the simulation was $N=512$. 
Inset: Plot of  $t(\lambda-\lambda_c)^{\nu_{||}}$ against $p(t)t^{\delta}$ using the previously obtained value of $\delta$.
The data collapse is obtained for $\nu_{||} \approx 2.0 \pm 0.1$.}
\label{nu-collapse-pc}
\end{figure}
From usual finite size scaling theory \cite{Hinrichsen:2000}, the order parameter is expected to follow a scaling relation of the form
\begin{equation}
\label{scaling}
X(t)\approx t^{-\delta}\mathcal{F}(t^{1/\nu_{||}}\Delta),
\end{equation}
where $\Delta=\lambda-\lambda_c$ and $\mathcal{F}$ is a universal scaling function of a form such that for large argument, 
the time dependence drops out ($\mathcal{F}(x)\sim x^{\delta \nu_{||}}$).
From the data collapse of the off-critical relaxation (Insets of Fig.~\ref{nu-collapse-o}), 
one can obtain the correlation time exponent to be $\nu_{||}\approx 1.5\pm0.1$ for $p$ and $\nu_{||}\approx 1.2\pm0.1$ for $O$. 

We carry out the above analysis for the fully connected version of the LCCC model. We only quote the main results of this version. 
The critical point is estimated to be $\lambda_c\approx 0.66659\pm0.00002$ for $N=1200$, 
slightly different from the lattice version. The critical exponent $\delta$ for $p$ turns out to be $1.2\pm0.1$, again close to 
value with the lattice measurement. The off-critical scaling for $p$ yields the correlation time exponent $\nu_{||}\approx 1.1\pm0.1$, 
slightly different from the lattice counterpart.  

We repeat the analysis for the fully connected version of model C.
For $N=512$, the critical point is found to be $0.500097 \pm 0.000001$, the same for the decay of
$O$ and $p$, close to the mean-field estimate (1/2).
We also get estimates of decay exponent $\delta$ as $0.500 \pm 0.005$ for $O$ and $0.521 \pm 0.005$ for $p$.
Using the above scaling relation for the data collapse (Eqn.(\ref{scaling}), we estimate values of exponent
$\nu_{||}$ as $0.10 \pm 0.01$ for $O$ and $2.00 \pm 0.02$ for $p$ (Fig.~\ref{nu-collapse-pc}).

The same analysis is repeated for the fully connected version of model G.
For $N=512$, the critical point is found to be $0.33338 \pm 0.00001$, the same for the decay of
$O$ and $p$, close to the mean-field estimate (1/3).
The estimates of the decay exponent $\delta$ are $0.585 \pm 0.001$ both for $O$ and $p$.
From the above scaling relation for the data collapse (Eqn.(\ref{scaling}), the estimates of exponent
$\nu_{||}$ are $1.6\pm0.1$ for $O$  and $\nu_{||} = 2.0 \pm 0.1$ for $p$. It is to
be noted that these exponents violate the scaling laws.
\begin{table*}
\caption{Table comparing the different quantities for the 3 models}
\centering
\begin{tabular}{|c|c|c|c|c|c|c|}
\hline

Model & $\lambda_c$ (Mean field) & Measured quantity & $\beta$ & $z$ & $\delta$ & $\nu_{||}$ \\
\hline
\hline
\multirow{2}{*}{LCCC} & $2/3$ & $O$ & $0.10(1)$ & $0.97(1)$ & $1.00(5)$ & $1.2(1)$\\
\cline{3-7}
                       &        & $p$  & $0.95(2)$ & $1.1(1)$  & $1.2(1)$ & $1.1(1)$ \\
\hline
\hline
\multirow{2}{*}{C} & $1/2$ & $O$ & $0.17(1)$ & $1.58(1)$ & $0.500(5)$ & $0.10(1)$\\
\cline{3-7}
                       &        & $p$  & $0.98(2)$ & $1.34(1)$  & $0.521(5)$ & $2.00(2)$ \\
\hline
\hline
\multirow{2}{*}{G} & $1/3$ & $O$ & $0.081(1)$ & $1.2(1)$ & $0.585(1)$ & $1.6(1)$\\
\cline{3-7}
                       &        & $p$  & $0.85(1)$ & $1.75(1)$  & $0.585(1)$ & $2.0(1)$ \\
\hline
\end{tabular}
\label{table:sum}
\end{table*} 
\begin{figure}
\begin{center}
\includegraphics[width=6.0cm,angle=270]{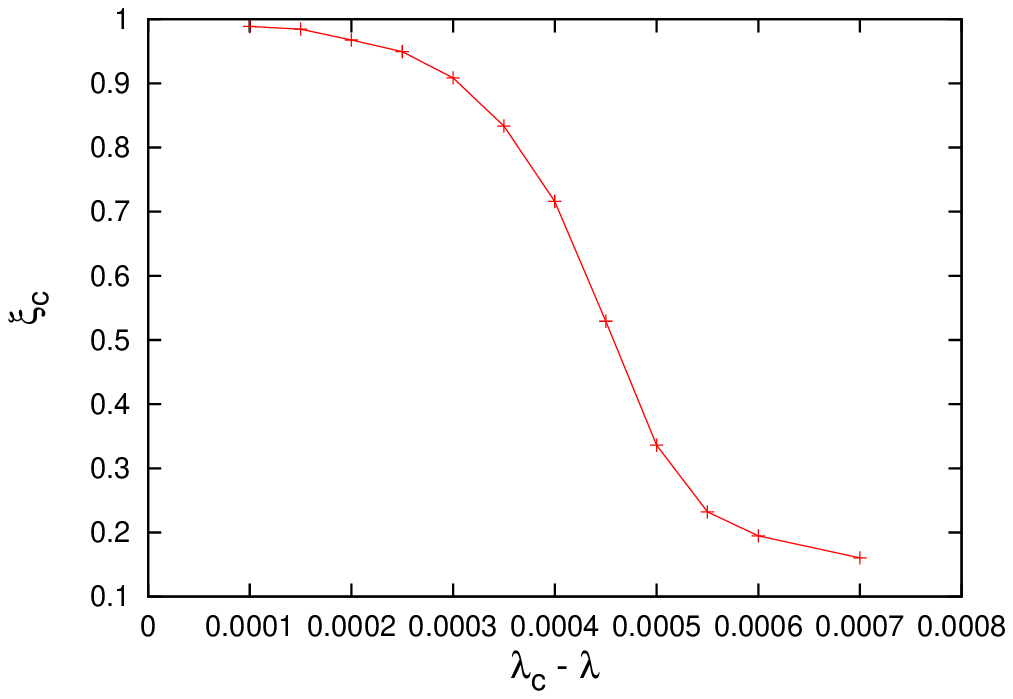}
   \caption{The correlation as a function of $(\lambda_c-\lambda)$
in $1d ~LCCC$ model for $N = 1000$.}
\end{center}
\label{fig:corr}
\end{figure}

In the quest of finding if there is a growing length scale in the system 
as the critical point is approached, we adopted the following strategy:
we begin with all agents having opinion $0.0$ except a particular agent was fixed at 
the negative extremity ($-1$) and was kept fixed forever in all subsequent updates.
Now as the system is allowed to relax, it is expected that the agent with the rigid 
opinion will have a neighborhood
of influence. This is a measure of the \textit{correlation length} $\xi_c = N_-/N$, 
where $N_-$ is the number of agents with opinion $o_i < -\epsilon$, where $\epsilon$
is a very small number (in our simulation we have taken $\epsilon = 0.005$). 
This is expected to grow as the critical point is approached from the sub-critical 
regime, which is indeed seen from the Monte Carlo simulations and will remain one
in the super-critical regime. In Fig.6 we have shown the growth of length scale
for the $1d ~LCCC$ model.

\section{A generalised model}
\label{sec:Sen}
\begin{figure}
\centering \includegraphics[width=6.0cm]{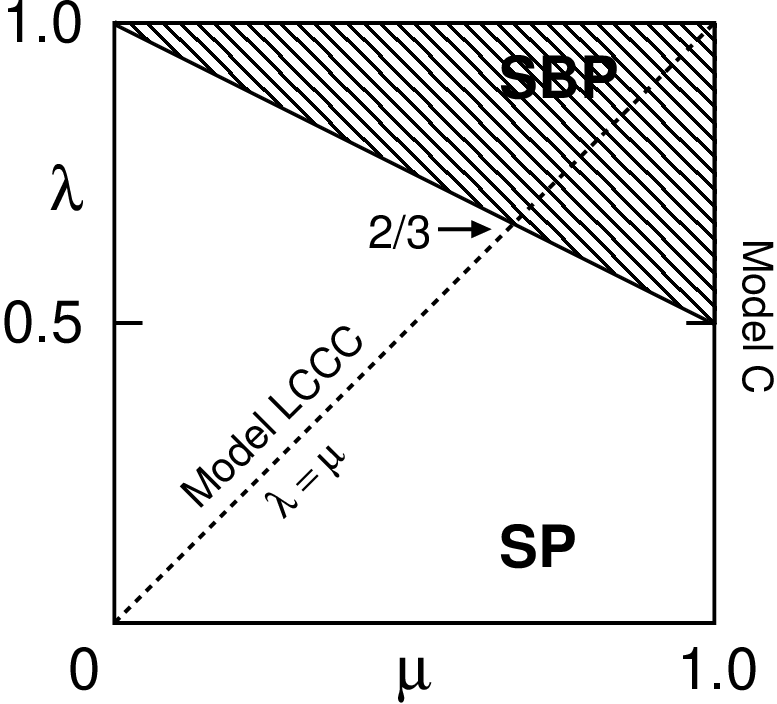}
   \caption{The schematic phase diagram of the generalised model. Distinct symmetric phase (SP)
and symmetry broken phase (SBP) exist. The $\lambda = \mu$ line corresponds to the Model LCCC
while $\mu=1$ line corresponds to Model C.}
\label{fig:delta}
\end{figure}
A generalised version of the LCCC model was introduced by Sen in Ref. 
\cite{Sen:2010} where the influencing parameter of the $i$th agent $\lambda_i$ 
is in general different from the conviction parameter $\mu_i$. Here the
binary opinion exchange is mathematically represented by
\begin{eqnarray}
\label{eq:glccc}
 o_i(t+1) &=& \lambda_i o_i(t) + \epsilon \mu_j o_j(t) \nonumber \\
 o_j(t+1) &=& \lambda_j o_j(t) + \epsilon^\prime \mu_i o_i(t)
\end{eqnarray} 
where the variables are as defined in previous sections. In the study
$\lambda$ and $\mu$ were assumed to be homogeneous. The special case
of $\lambda = \mu$ is the LCCC model. 
The limiting case of $\mu=1$ corresponds to the model C (Fig.~\ref{fig:delta}).
In the generalised case the
transition from symmetric to symmetry broken phase is determined by
both $\lambda$ and $\mu$, the mean field phase boundary given by $\lambda = 1 -\mu/2$.

\subsection{Non-universal behavior}
Along the phase boundary mentioned above, the critical behavior is reported to be strongly non-universal. 
Here also the relaxation time diverges close to the transition points along
the phase boundary as $\tau \sim |\lambda-\lambda_c|^{-z}$ for corresponding
values of $\mu_c$. But in this case $z$ varies with $\mu_c$ rather systematically, indicating a
non-universal behavior.  For $\mu_c=0.4$, $z=1.04\pm 0.01$; $\mu_c=2/3$, $z=1.10\pm 0.03$; $\mu_c=0.9$, $z=1.21\pm 0.01$. This non-universal behavior is also present in the order parameter exponent $\beta$. For the order parameter $O$, $\beta=0.079\pm 0.001$ for $\mu_c=0.4$ and $\beta=0.155\pm 0.001$ for $\mu_c=0.9$. 

The condensation fraction $p$ also shows similar
behaviour as $O$. The relaxation time diverges close to the transition points 
along
the phase boundary as $\tau_p \sim |\lambda-\lambda_c|^{-z_p}$ for 
corresponding values of $\mu_c$. The value of $z_p$ now only weakly varies with 
$\mu_c$ but is very close to $z$, indicating the existence of only one
time scale.
Also $p \sim (\lambda-\lambda_c)^{\beta_p}$ with $\beta_p\approx 0.91$ for $\mu_c=0.4$, $\approx 0.95$ for $\mu_c=2/3$ and $\approx 1.0$ for $\mu_c=0.9$. Here also, although the non-universality is present, it is very weak.
\section{A generalized map}
\label{sec:map}
Recently, an interesting map has been proposed by Chakrabarti in Ref.~\cite{Chakrabartimap}, limiting
cases of which correspond to known kinetic models. It has the following form:
\begin{equation}
\label{eq:ascmap}
 x(t+1) = \min \{(\alpha_1 + \epsilon_t \alpha_2)x(t) + \xi_t \alpha_3^n, \theta \}
\end{equation}
where $\alpha_1$, $\alpha_2$ and $\alpha_3$ are linear functions of a single parameter
$\alpha$,  ($0 \le \alpha_i \le  1$ for $i=1,2,3$),
with $-\infty \le n \le 1$ and $\theta$ taking
value either a positive, finite value (take $\theta = 1$, for convenience) or 
$\infty$ (sufficiently large value) and $\epsilon_t, \xi_t$ are uniform in $[0, 1]$ 
and are independent.

\begin{enumerate}
 \item[(a)]
When $\alpha_1 = \alpha$ and $\alpha_2 = \alpha_3 = 1-\alpha$, $n=1$, $\theta=\infty$,
$x(t+1) = (\alpha + \epsilon (1-\alpha))x(t) + \xi (1-\alpha)$ for which the steady
state distributions of $x$ are close to Gamma distributions for large $alpha$.
For $\xi=\epsilon$, this reduces to
$x(t+1) = \alpha x(t) + \epsilon (1-\alpha)(x(t) + 1)$, closely representing the CC model.

\item[(b)]
When $\alpha_1 = \alpha$ and $\alpha_2 = \alpha_3 = 1-\alpha$, $-\infty \le 0$, $\theta=\infty$,
then $x(t + 1) = {\alpha + \epsilon(1 - \alpha)}x(t) + \xi(1 - \alpha)^n$.
It is observed that $x$ is distributed as a power law as $P(x) = x^{-\frac{n-2}{n-1}}$.

\item[(c)]
When $\alpha_1=\alpha_2=\alpha$, $\alpha_3=0$, $n=1$ and $\theta=1$, then Eqn.~(\ref{eq:ascmap})
reduces to $x(t+1) = \min {\alpha(1+\epsilon)x(t), 1}$, which is nothing but the map version
of the LCCC model as given in Eqn.~(\ref{eq:maplccc}).
\end{enumerate}

\section{A model with preference}
\label{sec:new}
Agents are endowed with opinion $o_i$ to begin with, with $o_i$
distributed random uniformly in $[-1,+1]$. The society is also parametrized
by a conviction $\lambda$ (as in LCCC).

The interactions are binary. A pair of agents $i$ and $j$ meet and finds out who has the stronger
opinion (compare values of $o_i$ and $o_j$). If $o_i > o_j$, then
\begin{equation}
o_j = \lambda (o_j + \epsilon o_i),
\end{equation}
and if $o_i < o_j$
\begin{equation}
o_i = \lambda (o_i + \epsilon o_j),
\end{equation}
$\epsilon$ being the usual random number drawn from an uniform distribution in $[0,1]$.
Note that there is no dynamics if they have the same opinion (they agree).

\begin{figure}
 \centering \includegraphics[width=9.0cm]{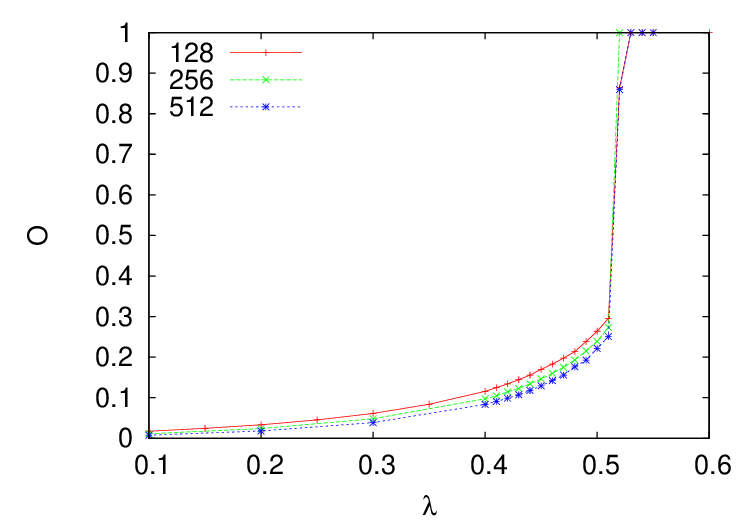}
\caption{Behavior or order parameter $O$ for different system sizes.}
\label{fig:newmodel}
\end{figure}

One computes the usual order parameter $O = |\sum_i o_i |/N$ and also the fraction $p$ of agents at $\pm 1$.
$O$ seems to undergo a phase transition at some $\lambda_c \simeq 0.52$ (Fig.~\ref{fig:newmodel})
while $p$ is a step function at the same $\lambda_c$. These signatures tempt us
to conclude that the phase transition here is discontinuous in nature.

\section{Summary  and Discussions}
\label{sec:sum}
We have reviewed here the studies on kinetic exchange model for opinion 
formation~\cite{Lallouache:2010a,Lallouache:2010b}, its two variants (proposed here) and a generalized
model~\cite{Sen:2010}.
In particular, we report the results from the study using non-equilibrium relaxation 
simulation technique to study the relaxation behavior of the order parameters for similar models. 

Here the lattice version of the LCCC model was proposed. 
The critical points and the critical exponents $\delta$ and $\nu_{||}$ for $1d$ version 
were found using two quantities, i.e., the average opinion $O$ (order parameter) and the condensation fraction $p$. 
The critical points for both the quantities are exactly the same, however, the values of the exponents differ slightly. The mean field version gave same critical point and slightly different exponent values. 
Then a simpler version of the LCCC model was studied. 
And finally we also study the effect of the feedback of the global opinion
formation in this model. 
A mean field calculation is made for both these variants, giving the critical points to be $1/2$ and $1/3$
respectively. These are again numerically verified within good accuracy.
Analysing the NER functions of the order parameters, the dynamical critical exponent, 
the order parameter exponent were estimated. 

Next, we discuss the generalized model with `influence' and `conviction'~\cite{Sen:2010}, and comment
on the limiting cases which correspond to models LCCC and C.
We also discuss the recently proposed map model, the special cases of which
resemble some known kinetic exchange models in literature.
The phase transitions observed in all the above mentioned cases has the unique feature
that the disordered phase
consists of individual opinions with values very near to $0$. While in other phase 
transitions
in the disordered phase the corresponding individual order parameters have finite
positive or negative values but collectively resulting to a value of nearly $0$.
We also propose a modified model in the same spirit as the models discussed earlier,
the only difference being that the agents are only influential over the ones with weaker opinion.
However, agents who agree, do not influence each other. The model shows clear characteristics
of a first order transition.

\section*{Acknowledgments}
The authors thank  P. Sen and M. Marsili for discussions.
B.K.C. acknowledges collaborations with A. S. Chakrabarti, A. Chakraborti, M. Lallouache.



\end{document}